# Optical gaps, mode patterns and dipole radiation in two-dimensional aperiodic photonic structures

Svetlana V. Boriskina, Ashwin Gopinath, and Luca Dal Negro

*Department of Electrical & Computer Engineering, Boston University,*
*8 Saint Mary's Street, Boston, Massachusetts 02215*

**Abstract:**
Based on the rigorous generalized Mie theory solution of Maxwell's equations for dielectric cylinders we theoretically investigate the optical properties of two-dimensional deterministic structures based on the Fibonacci, Thue-Morse and Rudin-Shapiro aperiodic sequences. In particular, we investigate band-gap formation and mode localization properties in aperiodic photonic structures based on the accurate calculation of their Local Density of States (LDOS). In addition, we explore the potential of photonic structures based on aperiodic order for the engineering of radiative rates and emission patterns in Erbium-doped silicon-rich nitride photonic structures.

**Keywords**: Photonic crystals, deterministic aperiodic photonic structures, local density of states, radiative rate enhancement, emission directionality

**PACS classification codes:** 42.25.Bs, 42.25.Fx, 42.55.Tv, 42.70.Qs

## 1. Introduction

Long-range ordered aperiodic photonic structures without translational symmetry offer an almost unexplored potential for the control and manipulation of localized field states. This novel class of photonic structures share distinctive physical properties with both periodic media, i.e. the formation of well-defined energy gaps, and disordered random media, i.e. the presence of localized eigenstates with high field enhancement and Q-factors. In particular, photonic quasi-crystals and deterministic aperiodic structures can lead to novel design schemes for light emitting devices based on the excitation of critically localized optical modes with unique transport properties [1,2]. Critical modes are spatially localized field states characteristic of deterministic aperiodic environments that possess fascinating scaling, spectral and localization properties [3]. In contrast with the Anderson localized modes of fully disordered media, critically localized states decay weaker than exponentially, most likely by a power law, and possess a rich self-similar structure [4]. While long-range ordered aperiodic photonic structures offer a large flexibility for the design of optimized light emitting devices, the theoretical understanding of the complex mechanisms governing optical gaps and mode formation in aperiodic structures becomes increasingly more important. The formation of photonic bandgaps and the existence of quasi-localized light states have already been demonstrated for one-dimensional (1D) and two-dimensional (2D) aperiodic structures based on the Fibonacci and the Thue-Morse sequences [5-7]. However, to the best of our knowledge, a rigorous investigation of the bandgaps and optical modes in more complex types of 2D aperiodic structures has not been reported so far.

In this work we will discuss the design of 2D aperiodic structures based on symbolic sequences with fascinating spectral properties described by quasi-periodic, singular continuous and absolutely-continuous (flat) Fourier spectra such as the Fibonacci, Thue-Morse and Rudin-Shapiro photonic structures, respectively [8]. Our theoretical analysis is based on the rigorous generalized 2D Mie theory for multiple cylindrical scatterers [9,10]. The method is computationally very efficient and allows considering the contributions of high-order multipolar orders in the electromagnetic analysis. In addition, based on our approach, we can construct the 2D Green's functions for the different aperiodic structures and utilize them in order to compute the local density of states at any desired location within the structures: $\rho(\vec{r},\omega) = -2\omega/\pi c^2 \, \text{Im}\{G(\vec{r},\vec{r},\omega)\}$ [10]. The rigorous calculation of the LDOS allows discussing the nature of the photonic band structure, mode localization and position-dependent enhancement of the spontaneous emission rates in aperiodic structures.

## 2. Optical gaps formation and critical modes patterns

We consider 2D photonic structures generated by arranging $N$ identical dielectric rods according to the aperiodic Fibonacci, Thue-Morse and Rudin-Shapiro inflation rules [8]. The refractive index of the dielectric rods is chosen to be 2.3, coincident with the refractive index of the Er-doped silicon-rich nitride material recently demonstrated in a light-emitting photonic structure [11]. Figure 1 shows the LDOS calculated at the origin of a reference square periodic lattice and of the three types of aperiodic photonic structures versus the normalized frequency parameter $a/\lambda$, for two orthogonal polarizations ($a$ is the nearest-neighbor center-to-center separation and λ is the wavelength). Similarly to periodic photonic crystals [12], deterministic aperiodic photonic structures composed of rods favor the formation of TM bandgaps, and two TM bandgaps can be observed in the considered frequency range. We notice from Figure 1 that the average spectral positions of the TM bandgaps do not change from structure to structure, suggesting that the scattering properties of individual rods are mostly responsible for band-gap formation in aperiodic structures, as only local spatial correlations are possible [13]. Incidentally, we notice that the distinctive fluctuations observed in the LDOS of aperiodic structures for both TE and TM polarizations are of special interest for the realization of unique quantum optical experiments [14].

reduction of the lasing threshold at the critical mode frequencies.

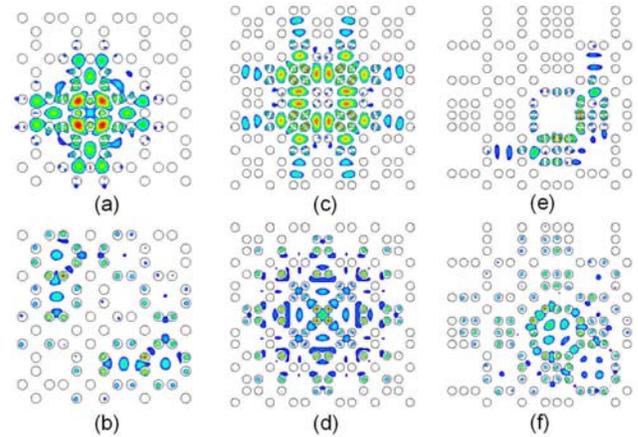

Fig. 2. Near-field intensity portraits of the TM- (a,c,e) and TE-polarized (b,d,f) high-Q optical modes supported by the aperiodic photonic structures considered in Fig. 1 with the lattice constant $a$=800 nm in the vicinity of λ=1.55 μm. (a): λ=1.538 μm, Q=198.4; (b): λ=1.526 μm, Q=191.1; (c): λ=1.502 μm, Q=257.0; (d): λ=1.516 μm, Q=537.0; (e): λ=1.544 μm, Q=201.2; (f): λ=1.52 μm, Q=323.7. For comparison, the Q-factor of the band-edge mode of the square-lattice photonic crystal with the resonant wavelength λ=1.626 μm is Q=133.8.

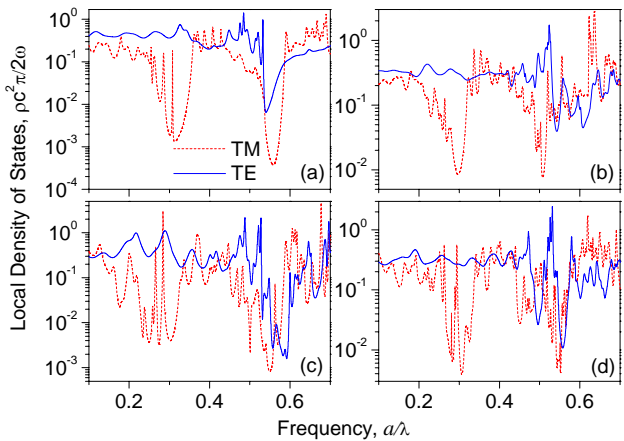

Fig. 1. The frequency-dependent LDOS for a line source located at the center of (a) square-lattice periodic ($N$=100), (b) Fibonacci ($N$=80), (c) Thue-Morse ($N$=128), and (d) Rudin-Shapiro ($N$=120) photonic structures composed of dielectric rods in air ($n_{rods}$=2.3, $r/a$=0.35).

Figure 1 also clearly demonstrates that high-Q modes, which are manifested as sharp peaks in the LDOS plots, are sustained in aperiodic photonic structures both inside the bandgap regions and near the band-edges. These sharp modes correspond to the critical states of the aperiodic structures, and their spectral density increases with increased structure's size [15]. The optical mode patterns corresponding to critical modes, shown in Figure 2, are generally very complex and may be either delocalized over the photonic structure or strongly localized in different areas of the structure. Furthermore, our results show that localized and quasi-localized high-Q modes can also appear outside of the bandgap region, similarly to the case of critical modes in Penrose-type quasi-crystals [16]. Owing to the high values of their Q-factors (which translate into longer photon lifetimes), critical modes in aperiodic photonic structures may be good candidates for the fabrication of low-threshold laser devices. It may also be expected that for critical modes of aperiodic structures the in-plane and out-of-plane quality factors are optimally balanced, similarly to the optimization of the optical confinement observed in a photonic crystal in the presence of structural disorder [17]. Such optimization of the in- and out-of-plane confinement of light may yield a

We also investigated the optical properties of 2D aperiodic photonic structures composed of cylindrical air-holes etched in a dielectric medium. Figure 3 shows the frequency-dependent LDOS calculated for a hexagonal periodic lattice and for Fibonacci, Thue-Morse and Rudin-Shapiro aperiodic structures. It can be observed that both periodic and aperiodic structures composed of air-holes favor the formation of TE bandgaps [12], whose spectral positions are determined by the array configuration rather than by the scattering properties of the individual air-holes [13]. In Figure 4 we show the near-field intensity distributions of the high-Q critical modes supported by the aperiodic structures with $a$=500 nm in the vicinity of 1.55 μm. Similarly to the case of dielectric rods, the band-edge mode of the hexagonal-lattice photonic crystal has a lower Q-factor (λ=1.496 μm, Q=13.8) than the ones of the critical modes supported by the aperiodic structures (shown in Figure 4).

### 3. Radiative rate and emission pattern manipulation

The high quality factors and the rich spectrum of critical modes characteristic of aperiodic photonic structures makes them very attractive platforms for the manipulation of radiative rates and emission patterns of embedded light emitting dipoles. We have shown if Figures 2 and 4 that localized and quasi-localized high-Q modes exhibit high values of LDOS at different spatial positions inside aperiodic structures. This behavior enables the possibility of strongly modifying the radiation properties of a light-emitting dipole located at these positions. According to the Fermi's golden rule [18,19], the decay rate of the emitter is proportional to LDOS intensity sampled at a specific location inside the structure (in vacuum, $\pi c^2/2\omega \, \rho_0(\vec{r},\omega) = 0.25$).

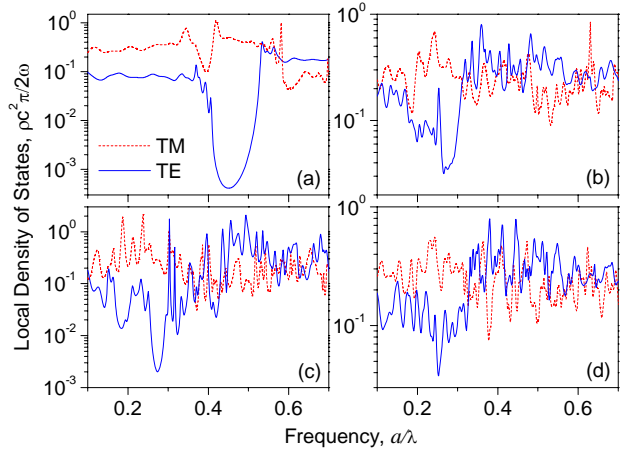

Fig. 3. The frequency-dependent LDOS for a line source located at the center of (a) triangular-lattice periodic (*N*=100), (b) Fibonacci (*N*=80), (c) Thue-Morse (*N*=128), and (d) Rudin-Shapiro (*N*=120) photonic structures composed of air-holes in dielectric ($n_{diel}$=2.3, $r/a$=0.45).

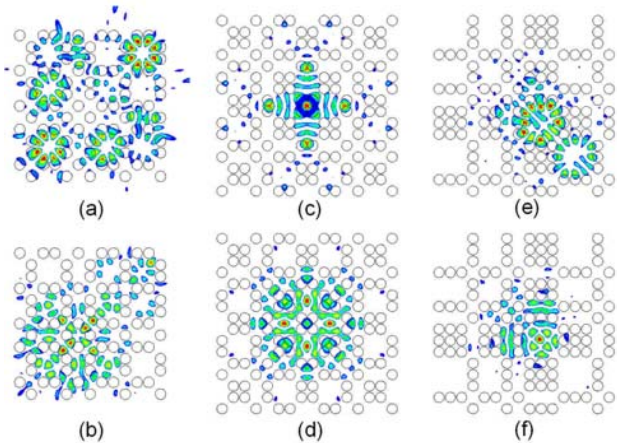

Fig. 4. Near-field intensity portraits of the TM- (a,c,e) and TE-polarized (b,d,f) high-Q optical modes supported by the aperiodic photonic structures considered in Fig. 3 with the lattice constant *a*=500 nm in the vicinity of λ=1.55 μm. (a): λ=1.408 μm, Q=52.7; (b): λ=1.489 μm, Q=60.75; (c): λ=1.655 μm, Q=74.6; (d): λ=1.581 μm, Q=331.0; (e): λ=1.465 μm, Q=117.0; (f): λ=1.511 μm, Q=118.6.

Considering the interesting case of Erbium ions as light-emitting dipoles, our calculations show that the λ=1.55 μm Er emission can be enhanced by a factor of 4.32 when Er ions are positioned at the center of the Thue-Morse structure, as shown in Fig. 4 (c) ($\rho/\rho_0 = 4.32$ at $x_0/a = 0$, $y_0/a = 0$). On the other hand, the stronger localization character of the modes of Rudin-Shapiro structures can provide larger radiative enhancement factors. By designing the localized modes of the Rudin-Shapiro structure, as shown in Fig. 4 (e), to be in resonance with the emission of Er ions located at the position $x_0/a = 2.35$, $y_0/a = -0.03$, we can obtained an overall emission enhancement of approximately 6 times with respect to its free-space value.

Finally, we study how the radiation mode patterns and directionality of an embedded source, which emits a cylindrical wave with a uniform radiation pattern in the free space, can be controlled and enhanced by the surrounding aperiodic photonic structure. Figure 5 shows the near- and far-field patterns calculated for a source embedded in the Fibonacci (a,b) and Rudin-Shapiro (c,d) photonic structures, respectively. In both cases, the TM-polarized line source is positioned at the center of the arrays ($x_0/a = 0$, $y_0/a = 0$), and radiates at the frequency coinciding with the resonant frequency of one of the critical eigenmodes of the structure (a,b: $a/\lambda = 0.6233$; c,d: $a/\lambda = 0.2897$). Directional single-beam radiation patterns can clearly be observed in Figures 5 (b) and 5 (d). Furthermore, the radiative rate of the source located at the center of the Fibonacci structure in Fig. 5 (a) is over an order of magnitude larger with respect to its free-space value ($\rho/\rho_0 = 10.25$).

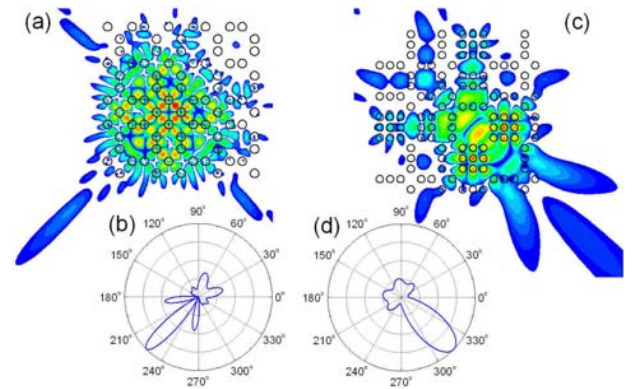

Fig. 5. Near field distributions (a,c) and the corresponding far-field radiation patterns (b,d) of the line source located at the center of the Fibonacci and Rudin-Shapiro photonic structures with the same parameters as in Fig. 1.

## 4. Conclusions

Based on a rigorous generalized Mie theory solution of Maxwell's equations we have explored the optical gap formation and mode localization properties of the three main types of deterministic aperiodic structures, namely Fibonacci, Thue-Morse and Rudin-Shapiro. We have demonstrated that strongly localized critical modes are formed in the bandgaps as well as in the band-edge regions of the LDOS spectra. Finally, by studying the interesting case of the 1.55 μm Erbium emission, we have demonstrated that up to a factor of 10 radiative enhancement with highly directional emission (beaming) can be achieved using the most-localized critical modes in the Rudin-Shapiro and Fibonacci structures.

### Acknowledgments

This work was partially supported by the Boston University College of Engineering Dean's Catalyst Award and the US Army Research Laboratory through the project: *Development of novel SERS substrates via "rationally" designed nanofabrication strategies.*